\documentclass[10pt,journal,compsoc,draftcls,onecolumn]{IEEEtran}
\usepackage{microtype}
\usepackage{graphicx}
\usepackage{subfigure}
\usepackage{booktabs} 
\usepackage{color}
\usepackage{amstext, amssymb, amsthm, amsmath,bm,rotating,array}
\usepackage{url}
\usepackage{epstopdf}
\usepackage{multirow}
\usepackage{multicol}
\usepackage{algorithm}
\usepackage{algorithmic}

\newtheorem{theorem}{Theorem}
\newtheorem{lemma}[theorem]{Lemma}

\newtheorem{assumption}{Assumption}

\newcommand{\vx}{{\bf x}}

\newcommand{\va}{{\bf a}}
\newcommand{\vb}{{\bf b}}

\newcommand{\vxi}{{\bm \xi}}
\newcommand{\valpha}{{\bm \alpha}}

\newcommand{\veta}{{\bm \eta}}

\newcommand{\ve}{{\bf e}}

\newcommand{\vz}{{\bf z}}
\newcommand{\vw}{{\bf w}}

\newcommand{\vu}{{\bf u}}

\newcommand{\vH}{{\bf H}}
\newcommand{\vE}{{\bf E}}

\newcommand{\vV}{{\bf V}}

\newcommand{\vY}{{\bf Y}}

\newcommand{\vX}{{\bf X}}

\newcommand{\vI}{{\bf I}}
\newcommand{\vK}{{\bf K}}
\newcommand{\vZ}{{\bf Z}}
\newcommand{\LShi}{\color{black}}

\ifCLASSOPTIONcompsoc
  \usepackage[nocompress]{cite}
\else
  \usepackage{cite}
\fi
\ifCLASSINFOpdf
\else
\fi
\begin{document}
%
\title{A Decentralized Framework for Kernel PCA with Projection Consensus Constraints}
%
%
%
%

\author{Fan~He,
        Ruikai~Yang,
        Lei~Shi$^*$,
        and~Xiaolin~Huang$^*$,~\IEEEmembership{Senior Member,~IEEE}
\IEEEcompsocitemizethanks{\IEEEcompsocthanksitem This work is supported by National Natural Science Foundation of China (No. 61977046), Shanghai Municipal Science and Technology Major Project (2021SHZDZX0102) and Shanghai Science and Technology Research Program (20JC1412700 and 19JC1420101).
\IEEEcompsocthanksitem F. He, R. Yang, and X. Huang are with Institute of Image Processing and Pattern Recognition, Shanghai Jiao Tong University, and the MOE Key Laboratory of System Control and Information Processing, 200240 Shanghai, P.R. China.\\
E-mail: \{hf-inspire, kai9906, xiaolinhuang\}@sjtu.edu.cn.
\IEEEcompsocthanksitem L. Shi is with Shanghai Key Laboratory for Contemporary Applied Mathematics, and also with School
of Mathematical Sciences, Fudan University, 200240 Shanghai, P. R. China.\\
E-mail: leishi@fudan.edu.cn.\\
X. Huang and L. Shi are the corresponding authors.
}
}

\IEEEtitleabstractindextext{%
\begin{abstract}
This paper studies kernel PCA in a decentralized setting, where data are distributively observed with full features in local nodes and a fusion center is prohibited.
Compared with linear PCA, the use of kernel brings challenges to the design of decentralized consensus optimization: the local projection directions are data-dependent. 
As a result, {\LShi the consensus constraint in distributed linear PCA} is no longer valid.
To overcome this problem, we propose a projection consensus constraint and obtain an effective decentralized consensus framework, where local solutions are expected to be the projection of the global solution on the column space of local dataset.
We also derive { a fully non-parametric, fast and convergent algorithm} based on alternative direction method of multiplier, of which each iteration is analytic and communication-efficient.
Experiments on a truly parallel architecture are conducted on real-world data, showing that the proposed decentralized algorithm is effective to utilize {\LShi information of other nodes} and takes great advantages in running time over the central kernel PCA.
\end{abstract}

}

\maketitle

\IEEEdisplaynontitleabstractindextext

%
\IEEEpeerreviewmaketitle

\IEEEraisesectionheading{\section{Introduction}\label{sec:introduction}}

\IEEEPARstart{P}{rincipal} Component Analysis (PCA) is a fundamental data processing technique with wide applications.
Given the data $\vX = [\vx_1\;\vx_2\;\cdots \;\vx_N]\in\mathcal{R}^{M\times N}$ {\LShi consisting of $N$ samples with each sample $x_i\in \mathcal{R}^M$ (viewed as a column vector)}, PCA considers the following {\LShi optimization problem
\begin{equation}\label{equ: linear pca}
    \begin{aligned}
        \min_{\vw\in\mathcal{R}^{M}} &\quad-\left\|\vw^\top\vX\right\|_2^2
        \quad \mathrm{s.t.} \; \|\vw\|_2^2= 1,
    \end{aligned}
\end{equation} aiming to find a direction $\vw$ such that the projection of data on this direction has maximal variance.} 
{\LShi Problem (\ref{equ: linear pca}) is equivalent to an eigendecomposition problem and the optimal solution $\vw^*$ is the unit eigenvector corresponding to the largest eigenvalue of the data covariance matrix $\vX\vX^\top$.}

Linear PCA problem has been well investigated in literature. {\LShi In this paper, we consider kernel PCA (kPCA, \cite{scholkopf1997kernel}), i.e., the nonlinear extension of PCA using the kernel trick. With the help of a nonlinear feature mapping $\phi:\mathcal{R}^M \rightarrow \mathcal{R}^P$ that maps the original data into a high-dimensional feature space $\mathcal{R}^P$, one can obtain nonlinear PCA via implementing linear PCA in the feature space. This process can be formulated as
\begin{equation}\label{equ: nonlinear-pca}
    \begin{aligned}
        \min_{\vw\in\mathcal{R}^P} &\quad-\left\|\vw^\top\phi(\vX)\right\|_2^2
        \quad \mathrm{s.t.} \; \|\vw\|_2^2= 1,
    \end{aligned}
\end{equation}
where we use $\phi(\vX) = [\phi(\vx_1) \; \phi(\vx_2) \;\cdots \;\phi(\vx_N)]\in \mathcal{R}^{P\times N}$ for brevity.} {\LShi The dimension of the feature space can be infinite, i.e., $P=\infty$. In this case, the feature space is taken to be the sequence space $\ell^2(\mathcal{R})$ consisting of all square summable sequences of real numbers, i.e., $\mathcal{R}^{\infty}:=\ell^2(\mathcal{R})=\left\{\vw=(w_i)_{i\geq 1} \subset \mathcal{R}: \sum_{i\geq 1} w_i^2 < \infty \right\}$. The sequence space $\ell^2(\mathcal{R})$, being equipped with inner product $\langle \va, \vb \rangle = \sum_{i\geq 1} a_ib_i$, for $\va=(a_i)_{i\geq 1}, \vb=(b_i)_{i\geq 1} \in \ell^2(\mathcal{R})$ is a Hilbert space with $\ell^2-$norm $\|\vw\|_2=\sqrt{\sum_{i\geq 1} \vw_i^2}$ for $\vw \in \ell^2(\mathcal{R})$. The Euclidean space $\mathcal{R}^P$, for $P\in \mathbb{N}$, can be regraded as a linear subspace of the sequence space $\ell^2(\mathcal{R})$. Hereinafter, with a slight abuse of notation, every element of $\mathcal{R}^P$ (whether $P\in \mathbb{N}$ or $P=\infty$)  will be viewed as a column vector, i.e., an $\mathcal{R}^{P\times 1}$ matrix, and the calculations among these vectors can follow the same rule from matrix operations. In particular, the inner product of $\va, \vb \in \mathcal{R}^P$ is $\va^\top \vb$, which is calculated through matrix multiplication. Define a positive definite kernel $\mathcal{K}:\mathcal{R}^M \times \mathcal{R}^M \rightarrow \mathcal{R}$ via $\mathcal{K}(\vx,\vx')=\phi(\vx)^\top \phi(\vx')$ that corresponds to the inner product of $\phi(\vx)$ and $\phi(\vx')$ in the feature space. Solving non-linear PCA problem (\ref{equ: nonlinear-pca}) can be reduced to seeking the eigenvector $\valpha\in \mathcal{R}^N$ with the largest eigenvalue $\lambda_1$ of the kernel matrix $\vK =[\mathcal{K}(\vx_i,\vx_j)]_{1 \leq i,j \leq N}\in \mathcal{R}^{N\times N}$. Then the solution of (\ref{equ: nonlinear-pca}) is given by $\vw^*=\phi(\vX)\valpha$. Here we require $\|\valpha\|_2=\frac{1}{\sqrt{\lambda_1}}$ as $\vw^*$ has to be normalized in feature space. In real data analysis, we usually choose a suitable positive definite kernel $\mathcal{K}$ before hand, or directly start with a data-dependent kernel matrix $\vK$. Under some mild conditions, e.g., the underlying reproducing kernel Hilbert space induced by $\mathcal{K}$ is separable, the above feature map $\phi$ always exists \cite{cucker2007learning, steinwart2008support}. For a new input $\vx$, the projection of $\phi(\vx')$ onto the direction $\vw^*$ is given by $(\vw^*)^\top \phi(\vx')=\valpha^\top \phi(\vX)^\top \phi(\vx')=\sum_{i=1}^N \alpha_i \mathcal{K}(\vx_i,\vx')$. Equipped with a kernel $\mathcal{K}$, kPCA solves the optimization problem (\ref{equ: nonlinear-pca}) and conduct nonlinear feature extraction, denoising and dimensionality reduction. The most fascinating part of kPCA is that we do not require the explicit form of $\phi$ during the entire computing procedure.}
However, in another sense, it also makes the design of variants of kPCA more challenging.

In recent years, distributed computing architectures are in high demand for dealing with large-scale or scattered {\LShi datasets}, which cannot be processed in a single node for various reasons as accelerating the {\LShi optimization processes}, protecting data privacy or reducing communication cost.
Thus, distributed algorithms {\LShi have attracted more and more interests due to the wide applications in industries and networked systems like wireless sensor networks \cite{SchizasA} and Internet of Things \cite{wang2020federated}.} Generally, distributed algorithms can be categorized {\LShi into} two groups: centralized and decentralized approaches.
Centralized algorithms require data or variables being aggregated in a fusion center (FC), which however, brings problems in many applications.
For example, when the number of nodes is very large, it is costly to reach the FC for each nodes with limited bandwidth. {Moreover, the entire centralized algorithm cannot work once the FC fails.}
By contrast, decentralized algorithms are more attractive because without FC, only information exchanges among neighborhood nodes are required and therefore it is more reliable and easier to {\LShi be extended} to large-scale cases.

Although highly desired, decentralized kernel {\LShi methods} have not been well developed. There are two fundamental difficulties. First, the solutions, {\LShi e.g.,} the regressor, the classifier, or the projection direction, in kernel methods are the combination of {\LShi basis functions $\mathcal{K}(\vx_i,\cdot)$ dependent on data}, { from which it follows that it is nearly impossible to require solutions in nodes are consistent}. Second, {without careful designing one may fail to get a fully non-parametric model, i.e., all primal variables are eliminated} by
the kernel trick that {\LShi directly evaluates the inner product in the feature space by the kernel.}




In this paper, we study the decentralized kPCA. To the best of our knowledge, there is only decentralized algorithm developed for linear PCA {\LShi in the existing literature.} In \cite{gang2019fast}, the decentralized linear PCA is modeled as follows,
{\LShi
\begin{equation}\label{equ: dlpca}
    \begin{aligned}
        \min_{\vw_j\in\mathcal{R}^{M}} &\quad-\sum_{j=1}^J\left\|\vw_j^\top\vX_j\right\|_2^2\\
        \mathrm{s.t.} &\quad \|\vw_j\|_2^2= 1,\; \forall j,\\
        & \quad       \vw_j = \vw_q, \forall q\in\Omega_j,\; \forall j,
    \end{aligned}
\end{equation}
where $\vX_j = [\vx^{(j)}_1\;\vx^{(j)}_2\; \cdots\; \vx^{(j)}_{N_j}]\in \mathcal{R}^{M\times N_j}$} and $\Omega_j$ are the data and the neighbors of node $j$, respectively. Clearly, its main idea is to add the consensus constraint $\vw_j = \vw_q$ to synchronize local solutions such that (\ref{equ: dlpca}) decouples across nodes.
But in kPCA, the local direction {\LShi $\vw_j$} is in {\LShi the column space of $\phi(\vX_j)=[\phi(\vx^{(j)}_1)\;\phi(\vx^{(j)}_2)\; \cdots\; \phi(\vx^{(j)}_{N_j})]\in \mathcal{R}^{P\times N_j}$. We denote this space by $\mathrm{span}\{\phi(\vX_j)\}$.} Directly {\LShi performing decentralized PCA (\ref{equ: dlpca}) in the feature space yields}

{\LShi
\begin{equation}\label{equ: hard model}
\begin{aligned}
    \min_{\vw_j} &\quad-\sum_{j=1}^J\left\|\vw_j^\top\phi(\vX_j)\right\|_2^2 \\
    \mathrm{s.t.}
    &\quad \vw_j \in \mathrm{span}\left\{\phi(\vX_j)\right\},\; \forall j,\\
    &\quad \vw_j = \vw_q, \forall q \in \Omega_j,\; \forall j, \\
    &\quad \|\vw_j\|_2^2 = 1,\forall j,
\end{aligned}
\end{equation}
which is, however, not working.} The essential problem is the representation discrepancy among different nodes, {\LShi which requires $\vw_j \in \bigcap_j \mathrm{span}\{\phi(\vX_j)\}$ for each $j$, while $\bigcap_j \mathrm{span}\{\phi(\vX_j)\}$ is very small or even empty such that the consensus constraints in (\ref{equ: hard model}) can not be satisfied.}

Therefore, in this paper we discard this approach.
Instead, {\LShi we propose a new consensus constraint based on the minimal distance between the global optimum and solutions in local nodes.}
Specifically, we introduce a global variable to link nodes and relax the consensus constraint to
{\LShi
$$\vw_j = \phi(\vX_j)\vK_j^{-1}\phi(\vX_j)^\top\vz, \forall j,$$}
where the global variable $\vz\in\cup_j \mathrm{span}\{\phi(\vX_j)\}$ {\LShi and
$$\vK_j =\left[\mathcal{K}(\vx^{(j)}_p,\vx^{(j)}_q)\right]_{1 \leq p,q \leq N_j}\in \mathcal{R}^{N_j\times N_j}.$$ The kernel matrix $\vK_j$ is invertible due to positive definiteness of the kernel $\mathcal{K}$.
}

In the following sections, we will prove that {\LShi these projection consensus constraints} will lead
the solution to be the projection of global optimum, i.e., {\LShi the solution of problem (\ref{equ: nonlinear-pca}) with the entire dataset}, on the {\LShi column}
space {\LShi $\mathrm{span}\{\phi(\vX_j)\}$ of local data}. More importantly, we will develop an efficient decentralized algorithm that is fully non-parametric: the whole solving procedure {\LShi only} involves {\LShi inner product $\phi(\vx)^\top\phi(\vx')$ that can be evaluated by $\mathcal{K}(\vx,\vx')$ directly, thus the primal variables $\vw_j, \vz$ and the feature map $\phi$ are eliminated when implementing the algorithm.}

We have the following contributions in this paper:
\begin{itemize}
    \item We propose a novel decentralized kPCA with {\LShi projection} consensus constraints. The local solution is proved to be the projection of the global optimum on the column space of local dataset.

    \item We design a fast solving algorithm based on the Alternative Direction Method of Multiplier (ADMM, \cite{boyd2011distributed}). In each iteration, the update is analytical, non-parametric, and communication-effective.


    \item The effectiveness of the proposed decentralized kPCA is verified by numerical experiments on a truly parallel architecture.
\end{itemize}

The rest of the paper is organized as the following.
In Section \ref{sec: relat}, {\LShi we review some related works}.
Section~\ref{sec: local} {\LShi formally introduces our decentralized framework for kPCA with projection consensus constraints} and theoretically discuss the solution propriety.
An efficient solving algorithm is designed in Section~\ref{sec: model}, and its convergence analysis is presented in Section \ref{sec: convergence}.
We then conduct numerical experiments in Section \ref{sec: exp}. {\LShi Finally, Section \ref{sec: conclusion} ends this paper with a brief conclusion}.

\section{Related works}
\label{sec: relat}
In the last decades, PCA and its variants, like kernel PCA, sparse PCA, online PCA, and robust PCA, have been deeply investigated \cite{scholkopf1997kernel}, \cite{jolliffe1986principal}, \cite{scholkopf1998nonlinear},  \cite{muller2001an}, \cite{vidal2005generalized}, \cite{zou2006sparse}, \cite{xu2012robust}, 
\cite{lloyd2014quantum}, \cite{huang2017indefinite} and successfully applied in many fields \cite{rosipal2001kernel}, \cite{zhang2010two},  \cite{gorban2008principal}.
Until very recently, there is still interesting progress.
For example, $\ell_1$-norm PCA and kernel PCA are discussed in \cite{markopoulos2017efficient}, \cite{tsagkarakis2018l1} and \cite{kim2020a}.
The convex formulation of kPCA is studied in \cite{alaiz2018convex}.
Truncated robust PCA model is newly proposed in \cite{nie2021truncated}.
PCA and robust PCA on tensor are studied in \cite{liu2018improved} and \cite{lu2020tensor}.


In recent years, distributed computing architecture is highly desired to deal with {\LShi big} data or distributed data.
For distributed data, there are two categories \cite{korada2011gossip}, \cite{wu2018a},  \cite{fan2019distributed}: feature-distributed and sample-distributed.
In feature-distributed setting, each node holds all samples but part of their features.
Several linear PCA algorithms are designed in this setting and are applied in wireless sensor network \cite{SchizasA}, \cite{BertrandDistributed}.
In this paper, we are interested in sample-distributed data, where each node holds part of the data with entire features.
In sample-distributed linear PCA, the key property is that \emph{the global covariance matrix is the sum of local covariance matrix}. 
Accordingly, one can design distributed algorithms on {\LShi an} FC to achieve high accuracy and efficient communication.
For example, distributed versions of power method and Oja method are proposed in \cite{wu2017the}. {\LShi The work}
\cite{ge2018minimax} designs distributed power methods for sparse PCA under differential privacy constraints.
Another differential private PCA algorithm is proposed in \cite{grammenos2020federated}, which is asynchronous and federated with a tree structure.
{\LShi For the sake of} communication efficiency, one-shot distributed algorithms are proposed in \cite{garber2017communication, fan2019distributed}, where rigorous statistical error analysis is established in the latter.
Other works on
distributed linear PCA can be found in, e.g.,
\cite{morral2012asynchronous},  \cite{liang2014improved}, \cite{anaraki2014memory}, \cite{boutsidis2016optimal}, \cite{huang2020communication}.

However, all {\LShi these} methods are restricted to {\LShi the case of} linear PCA, since the key property mentioned above is no longer true for {\LShi its kernelized counterpart}.
Due to that, there is only few discussions on distributed kPCA. Related works can be found in \cite{zheng2005an}, \cite{wang2006kernel}, which are actually to deal with large scale problem but are not in our distributed setting.
Recently,  the work \cite{balcan2016communication} discusses a distributed framework of kPCA, where subspace embedding and adaptive sampling are used to generate a representative subset to perform kPCA. However, the estimation {\LShi of} global solution is still in the centralized setting with an FC.
Different from distributed kPCA, recently, there are many researches focusing on distributed kernel regression and kernel support vector machine, proposing advanced algorithms and rigorous statistical  analysis \cite{forero2010consensus}, \cite{zhang2015divide}, \cite{chouvardas2016a}, \cite{guo2017learning}, \cite{mucke2018parallelizing}, \cite{bouboulis2018online}, \cite{koppel2018decentralized}, \cite{xu2019coke},  
\cite{lin2020distributed}.
However, many of them {\LShi still need an FC}, e.g., \cite{zhang2015divide}, \cite{guo2017learning}, \cite{mucke2018parallelizing}, 
\cite{lin2020distributed}.
As we discussed above, {\LShi the framework of} centralized distributed computing requires fusion to produce new estimation, which brings extra problems in real {\LShi data processing}.

Among the recent decentralized frameworks for kernel {\LShi methods}, { diffusion-based \cite{chouvardas2016a},  \cite{bouboulis2018online} and consensus-based \cite{forero2010consensus}, \cite{koppel2018decentralized},\cite{xu2019coke} algorithms are two most efficient approaches}.
In this paper, we consider the consensus-based algorithm, where consensus constraints are used such that the local variables are forced to agree with neighbors' variables.
Consensus optimization is widely used for {\LShi decentralized} problems, e.g., 
the decentralized abstract optimization \cite{notarstefano2011distributed} and 
the decentralized random convex programs \cite{carlone2014distributed}.
And recently, consensus optimization is applied to deal with decentralized linear PCA \cite{gang2019fast}.
In the context of decentralized kernel {\LShi methods}, there are studies on {\LShi decentralized support vector machine} \cite{forero2010consensus} and decentralized online kernel learning \cite{koppel2018decentralized}.
And recently, decentralized algorithms with consensus constraints and random features are proposed in \cite{xu2019coke}, \cite{richards2020decentralised}.
By adding consensus constraints on the difference among neighbors' variables, the original optimization is decoupled such that it can be solved in a decentralized manner.
Generally, consensus optimization can be efficiently solved by primal dual methods \cite{terelius2011decentralized} or ADMM, the latter of which has been well investigated in \cite{boyd2011distributed,hong2016convergence}.

In this paper, we focus on the decentralized framework of kPCA problem, which {\LShi is essentially different from} that of kernel regression or {\LShi support vector machine}, e.g., \cite{forero2010consensus}, \cite{koppel2018decentralized},\cite{xu2019coke}, \cite{richards2020decentralised}, where they add consensus constraints directly on local decision variables.
Specifically, in the linear case of optimization problem (\ref{equ: center problem}) {\LShi (i.e., $\phi(\vx)=\vx$, see Section \ref{sec: local} for details)}, if nodes $j$ and $q$ are neighbors, we can add the consensus constraint $\vw_j=\vw_q$ as previous works to force the two nodes communicate with each other, bringing better { performance} to the final solutions.
But as discussed previously, simply forcing $\vw_j=\vw_q$ is ineffective for kPCA problem {because the solution $\vw$ is data-dependent.}
We will further discuss this problem and propose our projection consensus constraints to deal with it in the next section.

\section{The projection consensus constraints}
\label{sec: local}

\subsection{Problem Background and \LShi Notations}
In this paper, we consider the sample-distributed kPCA in decentralized setting.
{\LShi The entire data} $\vX$ is not stored together, { and not accessible.}
In practice, the data are distributed observed in $J$ nodes: $\vX=[\vX_1\;\vX_2\;\cdots\;\vX_j] $ and the $j$-th node has data $\vX_j = [\vx^{(j)}_1\;\vx^{(j)}_2\; \cdots\; \vx^{(j)}_{N_j}]\in \mathcal{R}^{M\times N_j}$, {\LShi where $\sum_{j=1}^J N_j = N$.}
Suppose the neighborhood relation can be described by a symmetric, undirected network $\mathcal{G}=(\mathcal{V},\mathcal{E})$, where $\mathcal{V}$ and $\mathcal{E}$ are the set of nodes and edges, respectively.
We use $\Omega_j\subset \mathcal{V}$ to denote the neighbors of the $j$-th node, where $(l,j)\in \mathcal{E}, \forall l\in\Omega_j$.
A node in $\Omega_j$ could exchange data with node $j$ (but there may be noise).

{\LShi Throughout this paper, we require the feature map $\phi$ associated with a positive definite kernel $\mathcal{K}$ to be normalized in the feature space, i.e., $\phi(\vx)^\top\phi(\vx) = 1,\forall \vx \in \mathcal{R}^M$. Equivalently, we need $\mathcal{K}(\vx, \vx) = 1,\forall \vx \in \mathcal{R}^M$, which can be realized by normalizing the original kernel through $\frac{\mathcal{K}(\vx,\vx')}{\sqrt{\mathcal{K}(\vx,\vx)}\sqrt{\mathcal{K}(\vx',\vx')}}$. Recall that $\phi(\vX_j)=[\phi(\vx^{(j)}_1)\;\phi(\vx^{(j)}_2)\; \cdots\; \phi(\vx^{(j)}_{N_j})]\in \mathcal{R}^{P\times N_j}$. The kernel matrix $\vK_{(i,j)} \in \mathcal{R}^{N_i\times N_j}$ is defined as $\vK_{(i,j)} = \phi(\vX_i)^\top\phi(\vX_j)$,  where its $(p,q)$ entry is $\mathcal{K}(\vx^{(i)}_p,\vx^{(j)}_q)$. Here $\vx_p$ and $\vx_q$ are the $p$-th and $q$-th elements in $\vX_i$ and $\vX_j$, respectively. When $i=j$, we obtain the square kernel matrix $\vK_j=[\mathcal{K}(\vx^{(j)}_p,\vx^{(j)}_q)]_{1 \leq p,q \leq N_j}\in \mathcal{R}^{N_j\times N_j}$ defined on local data $\vX_j$.}

\begin{figure}[t]
\begin{center}
\centerline{\includegraphics[width=0.75\textwidth]{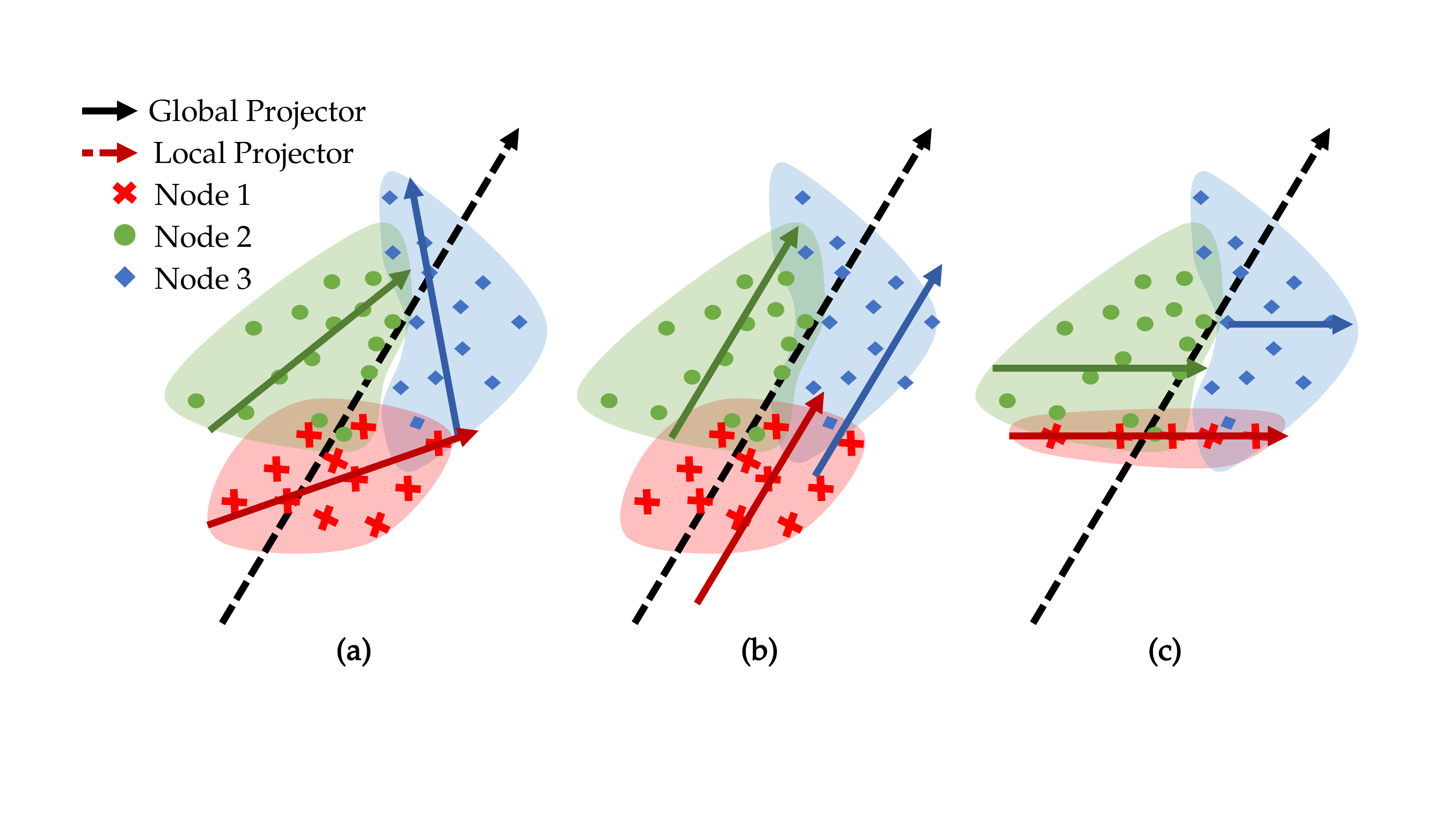}
}
\caption{A toy example of distributed dataset, showing that (a) there exists difference between the local solution (solid line) and the global solution (dotted line); (b) Applying the consensus constraint $\vw_1=\vw_2=\vw_3$, local solutions are all consistent with the global solution;
(c) In the extreme cases, the data of node $1$ (red star) all lie in a lower dimensional space (i.e., a line). Thus, the feasible domain is limited in a $1$-d space and applying the consensus constraint leads to poor solutions.}
\label{fig: problems}
\end{center}
\end{figure}

\subsection{Challenges in designing consensus constraints for kPCA.}
In the context of distributed dataset, kPCA actually faces two challenges. 
First is \textit{the data heterogeneous}, 
which is common in decentralized learning.
As shown by a toy example in Fig.~\ref{fig: problems} (a), the local optimum (solid lines) evaluated from local dataset are generally different from the global optimum (dotted line).
Thus, without an FC, each node needs to make the most of neighbors' information.
Fig.~\ref{fig: problems} (b) shows that the local solutions and the global solution are consistent if a consensus constraint $\vw_1 = \vw_2 = \vw_3$ is applied.
Second is special for decentralized kernel learning: \textit{the  representation  discrepancy}, just as we discussed in the introduction.
Consequently, simply forcing $\vw_p = \vw_q$  {\LShi is too strict to require as a consensus constraint. Sometimes such constraint may lead to very bad solutions. 
because its performance always dependents on the worst nodes.
}
An extreme example is shown in Fig.~\ref{fig: problems} (c), where the global dataset is $2$-dimensional but all data stored in node $1$ lie in a line.
Recall $\vw$ is data-dependent, and thus the feasible domain of $\vw_1$ is $1$-dimensional as well.
If we still pursue $\vw_1 = \vw_2 = \vw_3$ in this case, then the algorithm leads to a poor solution due to the limited feasible domain.
These two challenges lead to a dilemma: a good consensus optimization should make full use of others' information but avoid being affected by bad nodes.
To overcome this dilemma and design an effective consensus optimization for kPCA, next we will evaluate what is the ``best'' solution in local nodes.

\subsection{What should be pursued in local?}

As we discussed above, { the primal variables $\vw$ is data-dependent in kPCA.}
As a result, it is {\LShi very likely that the global solution cannot be represented by local data.}
That is, in decentralized setting, the local optimal decision variables are not always equal to the global optimum. Then a natural question is ``{\LShi what is the best solution we are pursuing at each node''.} Thus, before we present the decentralized algorithm, we need to {\LShi specify} the local optimal solution.
In local {\LShi node} $j$, we {\LShi expect a candidate local solution $\vw$ to satisfy} (i) $\vw$ is in {\LShi the column space $\mathrm{span}\{\phi(\vX_j)\}$}; (ii) {\LShi projection $\phi(\vX)$ on $\vw$ has the largest variance.}
Mathematically, we {\LShi incorporate the above two conditions into the following model,
\begin{equation}\label{equ: center problem}
    \begin{aligned}
        \tilde{\vw}_j =\mathop{\arg\min}\limits_{\vw\in \mathcal{R}^{P}} &\;-\left\|\vw^\top\phi(\vX)\right\|_2^2 \\
        \mathrm{s.t.}
        &\;\vw \in \mathrm{span}\{\phi(\vX_j)\},~ \|\vw\|_2^2 = 1.
    \end{aligned}
\end{equation}}

It is obvious that the solution of (\ref{equ: center problem}) is worse than that of (\ref{equ: nonlinear-pca}) because the feasible domain is shrunk.
Moreover, \textit{the optimal solution of different nodes are not the same} for the same reason (they have different feasible domain).
Note that (\ref{equ: center problem}) is unsolvable in primal space due to the implicit $\phi(\cdot)$.
By applying kernel trick again, the optimal solution of (\ref{equ: center problem}) is $\tilde{\vw}_j = \phi(\vX_j)\tilde{\valpha}_j$, where the actual variable $\tilde{\valpha}_j$ {\LShi is the eigenvector of $\phi(\vX_j)^\top\phi(\vX)\phi(\vX)^\top\phi(\vX_j)$ associated with the largest eigenvalue.} {\LShi Solving $\tilde{\valpha}_j$ can be
formulated as a generalized eigendecomposition problem, which in general lacks an explicit expression of $\tilde{\valpha}_j$.} {\LShi Moreover}, since the global data $\vX$ is unobtainable, computing $\tilde{\valpha_j}$ in local is impractical. And without {\LShi explicit} expressions, the relationship between $\tilde{\valpha}_j, \forall j$ is undefined.
As a result, we cannot design consensus constraints from this point of view.



\subsection{Decentralized framework with projection consensus constraints}
In this paper, we {\LShi adopt an alternative strategy}.
That is, we turn to minimize the distance between local solutions and the global solution, i.e.,
\begin{equation}\label{equ: local problem}
    \begin{aligned}
        \vw_j =\mathop{\arg\min}\limits_{\vw\in \mathcal{R}^{P}} &\;\|\vw-\vu\|_2^2 \\
        \mathrm{s.t.}
        &\;\vw \in \mathrm{span}\{\phi(\vX_j)\},\\
    \end{aligned}
\end{equation}
where $\vu$ is {\LShi the optimum of problem (\ref{equ: nonlinear-pca})}.
Then we can propose a new projection consensus constraint as
$$\vw_j = \phi(\vX_j)\vK_j^{-1}\phi(\vX_j)^\top\vu, \forall j = 1,\cdots,J,$$ {\LShi which is the solution of optimization problem (\ref{equ: local problem})}.
{\LShi In fact,} the local solution $\vw_j$ is the projection of the global solution $\vu$ on the column space of local dataset.
We now consider the following consensus optimization for kPCA,
\begin{equation}\label{equ: soft model}
\begin{aligned}
    \min_{\vw_j, \vz_j} &\quad-\sum_{j=1}^J\left\|\vw_j^\top\phi(\vX_j)\right\|_2^2 \\
    \mathrm{s.t.}
    &\quad \vw_j =  \phi(\vX_j)\vK_j^{-1} \phi(\vX_j)^\top\vz_j,\;\forall j = 1,\cdots, J \\
    &\quad \vz_j=\vz_q, \forall q \in \Omega_j,\; \forall j = 1,\cdots, J,\\
    &\quad \|\vz_j\|_2^2 = 1,\forall j = 1,\cdots, J.
\end{aligned}
\end{equation}
where each local machine only holds consensus constraints with nodes belonging to $\Omega_j$.
Different from (\ref{equ: hard model}), we introduce a new variable $\vz_j$ and pursue the consistency of it.
Actually, $\vz_j, \forall j$ are estimations of the global solution $\vu\in\mathcal{R}^{P}$ and thus the constraints $\vz_j=\vz_q, \forall q \in \Omega_j$ are reasonable.
As discussed before, we expect $\vw_j$ to be the optimal approximation of $\vu$ on the its feasible domain, i.e., the column space of $\phi(\vX_j)$.
Thus, the $\vw_j$ and $\vz_j$ are associated by the constraint $\vw_j =  \phi(\vX_j)\vK_j^{-1} \phi(\vX_j)^\top\vz_j$.

Before presenting the solving algorithm in the next section, we first evaluate the effectiveness of model (\ref{equ: soft model}),  where we need a trivial assumption on the network topology.
\begin{assumption}\label{asmp: cycle}
The undirected graph $\mathcal{G} = (\mathcal{V},\mathcal{E})$ is connected.
\end{assumption}

{\LShi Under Assumption~\ref{asmp: cycle}, one can easily see that the problem (\ref{equ: soft model}) is minimized when achieving consensus at each nodes. We summarize this observation as the following theorem. }

\begin{theorem}\label{the: equivalence}
Suppose {\LShi that} Assumption~\ref{asmp: cycle} is satisfied.
If $\{\vw_j^*, \vz_j^*\}$, $\forall j =1,\cdots,J$ are the optimal solution of problem (\ref{equ: soft model}), then $\vz_1^*=\vz_2^*=\cdots=\vz_J^*=\vu$, where $\vu$ {\LShi is the solution of optimization problem (\ref{equ: nonlinear-pca})}.
\end{theorem}

\section{Decentralized algorithm for kPCA}
\label{sec: model}
Hereinbefore, we explain why we cannot directly extend existing frameworks of decentralized linear PCA and other kernel methods to decentralized kPCA.
And based on the optimal approximation, a projection consensus constraint is proposed with the corresponding optimization (\ref{equ: soft model}).
In this section, we will derive a fast solving algorithm for it. Notice in kPCA, we do not know and cannot access the nonlinear mapping $\phi$. Thus, we need carefully and exquisitely design the algorithm to obtain a fully non-parametric algorithm that does not calculate $\phi$, although it formally appears in (\ref{equ: soft model}).

\subsection{Algorithm}
If $\phi$ is known or can be accessed, consensus optimization (\ref{equ: soft model}) is relatively easy to solve. 
But in kernel learning, the implicit mapping $\phi$ makes the problem quite difficult.
Specifically, the difficulty lies in the consensus constraint $\vz_j=\vz_q$, which exists in the unknown feature space.
To handle the consensus constraint in a non-parametric way, \cite{forero2010consensus, koppel2018decentralized} propose to transfer the constraint into its necessary condition  $\phi(\vY)^\top\vz_j=\phi(\vY)^\top\vz_q$ instead, where $\vY$ may be local dataset $\vX_j$ \cite{koppel2018decentralized} or a public dataset \cite{forero2010consensus}.
In this paper, we follows this idea with modification and propose the form below, $$\phi(\vX_j)\valpha_j =  \phi(\vX_j)\vK_j^{-1} \phi(\vX_j)^\top\vz_q,$$ to approximate $\vz_j=\vz_q$.
The advantage is that the subproblem corresponding to $\vz_j$ is independent across $j=1,\cdots,J$ now.
As a result, we can construct an analytic expression of $\vz_j$ in every ADMM iteration.

Let us reformulate the consensus optimization.
Recalling that  $\phi: \mathcal{R}^M\rightarrow \mathcal{R}^P$, we define the following matrices:
\begin{itemize}
    \item     $\vZ = [\vz_1\; \vz_2\; \cdots\; \vz_J] \in \mathcal{R}^{P\times J}, \quad \vz\in\mathcal{R}^{P   }    $.
    In the following solving process, $\vZ$ will always be implicit such that $\phi$ is unnecessary to know in the final algorithm.
    \item
    $\ve_j = [{ 0\; }\cdots\; {0\; } 1 {\; 0\; }\cdots {0}]^\top \in \mathcal{R}^{J }$:
    $\ve_j$ consists $J$ elements, which are all zeros but the $j$-th one is $1$.
    Thus, we can extract $\vz_j$ by $\vZ \ve_j$.
    \item $\vxi_j = [\ve_{q_1}\;\ve_{q_2}\;\cdots]\in \mathcal{R}^{J \times |\Omega_j|}$:
    the $i$-th column of $\vxi_j$ is $\ve_{q_i}$, where $q_1, q_2,\cdots$ are elements of $\Omega_j$.
    Thus, the $j$-th machine can extract its neighbors' $\vz$ by $\vZ\vxi_j$.
    \item $\vE_j = [1,\; \cdots,\; 1]\in\mathcal{R}^{|\Omega_j|}$, which is a row vector.
\end{itemize}
By substituting $\vw_j = \phi(\vX_j)\valpha_j$ to (\ref{equ: soft model}) and using $\phi(\vX_j)\valpha_j =  \phi(\vX_j)\vK_j^{-1} \phi(\vX_j)^\top\vz_q$ to replace $\vz_j=\vz_q$, we reformulate the optimization problem as follows,
\begin{equation*}\label{equ: ADMM-hard}
\begin{aligned}
    \min_{\valpha_j, \vZ}\quad &-\sum_{j=1}^J\left\|\valpha_j^\top\vK_j\right\|_2^2\\
    \mathrm{s.t.}\quad &  \phi(\vX_j)\valpha_j\vE_j =  \phi(\vX_j)\vK_j^{-1} \phi(\vX_j)^\top\vZ\vxi_j, \forall j = 1,\cdots, J, \\
    & \|\vz_j\|^2_2 \leq 1, \forall j = 1,\cdots, J.
\end{aligned}
\end{equation*}
Here we also relax the constraint $\|\vz\|_2^2 = 1$ to $\|\vz\|_2^2 \leq 1$ such that the feasible domain is convex.
This is a common relaxation and the optimal solutions are proofed to remain the same.
Then, we first deal with the update of $\vZ$.
The augmented Lagrangian is,
\begin{equation}\label{equ: augmented lagrangian}
    \begin{aligned}
        &\mathcal{L}(\valpha_j, \vZ; \veta_j) = -\sum_{j=1}^J\left\|\valpha_j^\top\vK_j\right\|_2^2 
        +\mathrm{tr}\left(\sum_{j=1}^J\veta_j^\top( \phi(\vX_j)\valpha_j\vE_j- \phi(\vX_j)\vK_j^{-1} \phi(\vX_j)^\top\vZ\vxi_j)\right)\\
        &+\frac{\rho}{2}\sum_{j=1}^J\left(\left\| \phi(\vX_j)\valpha_j\vE_j- \phi(\vX_j)\vK_j^{-1} \phi(\vX_j)^\top\vZ\vxi_j\right\|_2^2\right) \\
        &\mathrm{s.t.}\quad \|\vz_j\|^2_2 \leq 1
    \end{aligned}
\end{equation}
Fixing $\valpha_j$ and $\veta_j$, the $\vZ$-problem is
\begin{equation*}
    \begin{aligned}
        &\min_{\vZ} =
        -\mathrm{tr}\left(\sum_{j=1}^J\veta_j^\top \phi(\vX_j)\vK_j^{-1} \phi(\vX_j)^\top\vZ\vxi_j\right)
        +\frac{\rho}{2}\sum_{j=1}^J\left(\left\| \phi(\vX_j)\valpha_j\vE_j- \phi(\vX_j)\vK_j^{-1} \phi(\vX_j)^\top\vZ\vxi_j\right\|_2^2\right)\\
        &\mathrm{s.t.}\quad \|\vz_j\|^2_2 \leq 1
    \end{aligned}
\end{equation*}
It is a quadratic constrained quadratic convex programming.
Take the derivative with respect to $\vZ$,
\begin{equation*}\label{equ: dL/dZ}
\begin{aligned}
    &\frac{\partial \mathcal{L}(\valpha_j, \vZ; \veta_j)}{\partial \vZ}
    =\rho\sum_{j=1}^J\left( \phi(\vX_j)\vK_j^{-1} \phi(\vX_j)^\top\vZ \vxi_j\vxi_j^\top\right) 
    -\sum_{j=1}^J\left( \phi(\vX_j)\vK_j^{-1} \phi(\vX_j)^\top\veta_j\vxi_j^\top + \rho \phi(\vX_j)\valpha_j \vE_j\vxi_j^\top\right)
\end{aligned}
\end{equation*}
However, since $ \phi(\vX_j)\vK_j^{-1} \phi(\vX_j)^\top$ is implicit, it is intractable to obtain the solution.
Thus, we consider a relaxation optimization problem as follows,
\begin{equation}\label{equ: U-problem}
    \begin{aligned}
        &\min_{\vZ}~\mathcal{U}(\valpha_j,\vZ;\veta_j)\\
        =& \min_{\vZ}-\mathrm{tr}\left(\sum_{j=1}^J\veta_j^\top \phi(\vX_j)\vK_j^{-1} \phi(\vX_j)^\top\vZ\vxi_j\right)
        +\frac{\rho}{2}\sum_{j=1}^J\left(\| \phi(\vX_j)\valpha_j\vE_j- \vZ\vxi_j\|_2^2\right)\\
        &\mathrm{s.t.}\quad \|\vz_j\|^2_2 \leq 1, \forall j.
    \end{aligned}
\end{equation}
This problem has an analytic solution and it is separable across $\vz_j, \forall j = 1,\cdots,J$.
That is, the update of $\vz_j$ is independent with others.

The derivative with respect to $\vZ$ is
\begin{equation*}\label{equ: dU/dZ}
\begin{aligned}
    &\frac{\partial \mathcal{U}(\valpha_j, \vZ; \veta_j)}{\partial \vZ}
    =\rho\vZ \sum_{j=1}^J\left(\vxi_j\vxi_j^\top\right) 
    -\sum_{j=1}^J \phi(\vX_j)\left(\vK_j^{-1} \phi(\vX_j)^\top\veta_j + \rho \valpha_j \vE_j\right)\vxi_j^\top
\end{aligned}
\end{equation*}
Setting this derivative to zero, we have,
\begin{equation*}\label{equ: opt z-1}
\begin{aligned}
    \hat{\vZ}^{(t)} &= \sum_{j=1}^J \phi(\vX_j)(\vK_j^{-1} \phi(\vX_j)^\top\veta_j^{(t)} + \rho \valpha_j^{(t)}\vE_j)\vxi_j^\top\vH.
\end{aligned}
\end{equation*}
where $$\vH=\left[\rho\sum_{j=1}^J (\vxi_j\vxi_j^\top)\right]^{-1} = \mathrm{diag}\left\{\frac{1}{\rho|\Omega_1|},\;\cdots,\;\frac{1}{\rho|\Omega_J|}\right\},$$ 
and $|\Omega_j|$ is the number of neighbors of node $j$.
Therefore, we require every $\Omega_j$ should contain at least one element.
Then, we have
\begin{equation}\label{equ: ze}
\begin{aligned}
    &\hat{\vz}_j^{(t)} = \hat{\vZ}^{(t)}\ve_j \\
    &= \sum_{l=1}^J\phi(\vX_l)\left( \vK_l^{-1} \phi(\vX_l)^\top\veta_l^{(t)} + \rho \valpha_l^{(t)}\vE_l\right)\vxi_l^\top\vH\ve_j \\
    &= \sum_{l\in\Omega_j}\phi(\vX_l)\left( \vK_l^{-1} \phi(\vX_l)^\top\veta_l^{(t)}\vxi_l^\top\vH\ve_j  + \frac{1}{|\Omega_j|}\valpha_l^{(t)}\right).
\end{aligned}
\end{equation}

Now we can compute $\|\hat{\vz}_j\|_2^2 = \|\hat{\vZ}^{(t)}\ve_j\|_2^2$ in node $j$.
Because only $\phi(\cdot)^\top\phi(\cdot)$ is involved and node $j$ has all needed data $\vX_l, \forall l\in \Omega_j$, though there may be noise.
For the same reason, node $j$ can compute $\phi(\vX_l)^\top\vz_j , \forall l\in\Omega_j$, of which the analytic expression is as follows,
\begin{equation}\label{equ: phi_z}
\begin{aligned}
    &\phi(\vX_l)^\top\vz_j  
    = \left\{
    \begin{array}{ll}             \phi(\vX_l)^\top\hat{\vZ}^{(t)}\ve_j,& \; \mathrm{if\;} \|\hat{\vZ}^{(t)}\ve_j\|_2^2 \leq 1,\\      \phi(\vX_l)^\top\hat{\vZ}^{(t)}\ve_j/\|\hat{\vZ}^{(t)}\ve_j\|_2, &\;\mathrm{otherwise}.
    \end{array}
    \right.
\end{aligned}
\end{equation}

\begin{figure*}[t]
\vskip 0.2in
\begin{center}
\centerline{\includegraphics[width=0.8\textwidth]{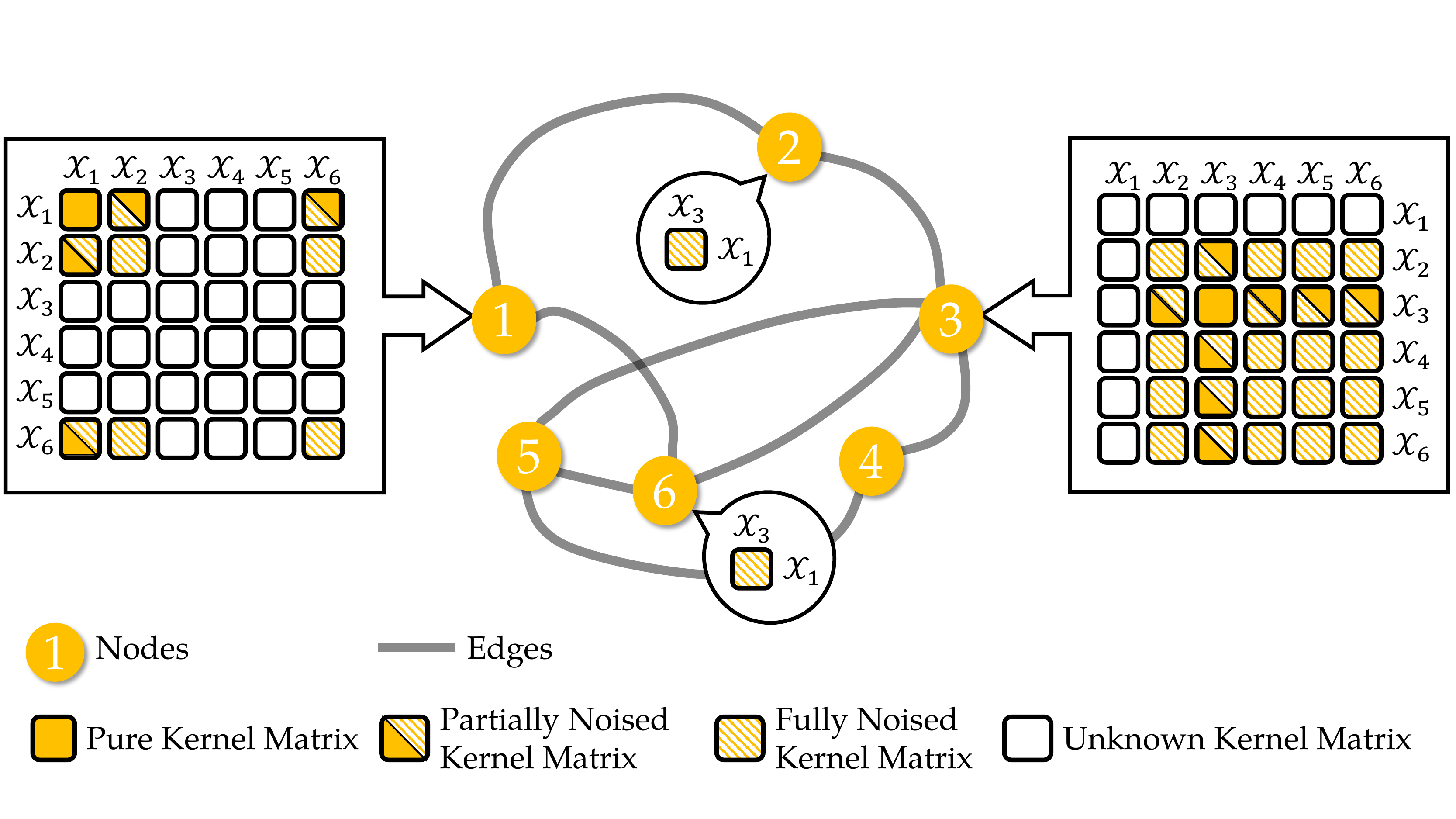}
}
\caption{A toy example of a network (middle) and the corresponding estimations of global kernel matrix in node $1$ and $3$ (left, right). Each node holds different information, so they have different estimations of the global kernel matrix.}
\label{fig: toy example}
\end{center}
\vskip -0.2in
\end{figure*}

The $\valpha_j$-problem is
\begin{equation*}
    \begin{aligned}
        &\min_{\valpha_j} \frac{\rho}{2}\left\| \phi(\vX_j)\valpha_j\vE_j\right\|_2^2 -\left\|\valpha_j^\top\vK_j\right\|_2^2   -\mathrm{tr}\left(\rho\vxi_j^\top\vZ^\top \phi(\vX_j)\valpha_j\vE_j\right) +\mathrm{tr}(\veta_j^\top \phi(\vX_j)\valpha_j\vE_j).
    \end{aligned}
\end{equation*}
This unconstrained quadratic programming is easy to solve.
Take the derivative of the objective function with respect to $\valpha_j$, and we have
\begin{equation*}\label{equ: dL/dalpha}
\begin{aligned}
    &\frac{\partial\mathcal{L}(\valpha_j, \vZ; \veta_j)}{\partial \valpha_j}
    = -2\vK_j^2\valpha_j  + \rho|\Omega_j|\vK_j\valpha_j +  \phi(\vX_j)^\top(\veta_j-\rho\vZ \vxi_j)\vE_j^\top,
\end{aligned}
\end{equation*}
where we use $\vK_j^2$ to denote $\vK_j\vK_j$.
Then $\valpha_j$ can be quickly computed by knowing the expression of $\phi(\vX_j)^\top\veta_j$ and $\phi(\vX_j)^\top\vZ \vxi_j$, i.e.,
\begin{equation}\label{equ: opt-alpha}
    \valpha_j = \left[\rho|\Omega_j|\vK_j-2\vK_j^2 \right]^{-1}\left(\phi(\vX_j)^\top(\rho\vZ \vxi_j-\veta_j)\vE_j^\top\right).
\end{equation}

Finally, the Lagrange multiplier matrices $\{\veta_j\}, \forall j$ are updated by the following gradient ascent iteration,
\begin{equation*}
    \begin{aligned}
        \veta_j^{(t+1)} &= \veta_j^{(t)} + \rho\phi(\vX_j)\left( \valpha_j\vE_j- \vK_j^{-1} \phi(\vX_j)^\top \vZ \vxi_j \right).
    \end{aligned}
\end{equation*}
Since $\phi(\cdot)$ is implicit, $\veta_j$ is unknown as well.
But fortunately, the updates for $\vZ$ and $\valpha_j$ both only involve $ \phi(\vX_j)^\top\veta_j$, which can be obtained by the following update,
\begin{equation}\label{equ: update  phi eta}
    \begin{aligned}
    \phi(\vX_j)^\top\veta_j^{(t+1)} &=  \phi(\vX_j)^\top\veta_j^{(t)} 
     + \rho\left( \vK_j\valpha_j\vE_j- \phi(\vX_j)^\top\vZ\vxi_j\right).
    \end{aligned}
\end{equation}
That is, we successfully constructed an algorithm which is fully non-parametric. 
Finally, the whole algorithm is summarized in Alg.~\ref{alg: soft}.

\begin{algorithm}[tb]
   \caption{ADMM-based Decentralized Algorithm of (\ref{equ: soft model}).}
   \label{alg: soft}
\begin{algorithmic}
   \STATE {\bfseries Input:} data $\vX_j$ and the constant matrices $\{\vxi_j\}_{j=1}^J, H$ related to the network topology.
   \STATE {\bfseries Output:} Vectors $\{\valpha_j\}_{j=1}^J$.
   \STATE \textbf{For all local nodes $j=1,\cdots,J$, do in parallel:}
   \STATE Set the ADMM hyper-parameter $\rho$.
   \STATE Randomly initialize $\valpha_j^{(0)}$. Initialize $\veta_j^{(0)} = 0$.
   \STATE Distributes $\vX_j$ to neighbors.
   \STATE Receive neighbors' $\vX_l$.
   Calculate $$\mathcal{K}(\vX_p, \vX_ q), \forall p,q\in\Omega_j.$$
   \STATE Set $t=0$.
   \REPEAT
   \STATE Distributes $\valpha_j^{(t)}$ and $\phi(\vX_j)^\top\veta_j^{(t)}\vxi_j\vH\ve_l$ to all the neighbors.
   \STATE Receive $\valpha_l^{(t)}$ and $\phi(\vX_l)^\top\veta_l^{(t)}\vxi_l\vH \ve_j$ from all the neighbors $l$.
   \STATE Solve the $\vZ$-problem:
   \begin{equation*}
       \begin{aligned}
           \vZ^{(t)} &= \arg \min_{\vZ} \mathcal{U}(\valpha_j^{(t)}, \vZ, \veta^{(t)})\\
           &\mathrm{s.t.}\;\|\vz_j\|_2^2\leq 1.
       \end{aligned}
   \end{equation*}
   according to (\ref{equ: ze}, \ref{equ: phi_z}), and then compute and distributes $\phi(\vX_l)^\top\vz_j^{(t)}$ to neighbors $l$.
   \STATE Receive $\phi(\vX_j)^\top\vz_l^{(t)}$ from all neighbors $l$.
   \STATE Solve the $\valpha_j$-problem according to (\ref{equ: opt-alpha}):
   \begin{equation*}
       \begin{aligned}
        \valpha_j^{(t+1)} &= \arg\min_{\valpha_j} \mathcal{L}(\valpha_j, {\vZ}^{(t)}, \veta^{(t)})
       \end{aligned}
   \end{equation*}
   \STATE Update
   \begin{equation*}
       \begin{aligned}
        &\phi(\vX_j)^\top\veta_j^{(t+1)} =  \phi(\vX_j)^\top\veta_j^{(t)}
        + \rho\left( \vK_j\valpha_j^{(t+1)}\vE_j- \phi(\vX_j)^\top\vZ^{(t)}\vxi_j\right)
       \end{aligned}
   \end{equation*}
   \STATE $t=t+1$.
   \UNTIL{the stop criteria is achieved.}
\end{algorithmic}
\end{algorithm}

\subsection{Discussion}
As a typical decentralized algorithm, Alg.~\ref{alg: soft} utilizes the information gap among nodes.
To illustrate the information fusion, we consider a network with $6$ nodes shown in Fig.~\ref{fig: toy example}. 
Supposing the noised raw data are exchanged among neighbors, we present the local estimations of global kernel matrix in node $1$ and $3$.
Obviously, node $3$ has many neighbors and has nearly full information but without that of node $1$.
Thus, local optimum of node $3$ still has room for improvement to better approximate the global optimum.
On the other side, node $1$ only has two neighbors, leading to little information and the poor performance of local solution.
By applying Alg.~\ref{alg: soft}, they implicitly communicate with each other through node $2$ and $6$.
Specifically, $\vK_{(1,3)}$ are computed in node $2$ and $6$ (with noise), of which the information is implicitly packaged in $\phi(\vX_1)^\top[\vz_2\;\vz_6]$ and $\phi(\vX_3)^\top[\vz_2\; \vz_6]$ and send to node $1$ and $3$, respectively.
By this way, Alg.~\ref{alg: soft} is effective to enhance the performance of local solutions $\vw_j$.

Alg.~\ref{alg: soft} is fast and fully non-parametric, leading to the practicability.
Actually, only three groups of variables are associated, i.e., $\valpha_j\in\mathcal{R}^{N_j },\; \phi(\vX_j)^\top\veta_j\in\mathcal{R}^{N_j\times |\Omega_j|},\;\forall j$ and $\phi(\vX_j)^\top\vz_q\in\mathcal{R}^{N_j }, \forall j,\; \forall q\in\Omega_j$.
All of them exist in dual space and can be computed via the kernel trick.
These variables are also used for communication, and thus we can analyze the communication cost of Alg.~\ref{alg: soft}.
In every ADMM iteration, node $j$ has two rounds communication:
\begin{itemize}
    \item Distribute $\phi(\vX_l)^\top\vz_j$ to and receive $\phi(\vX_j)^\top\vz_l$ from neighbor $l,\;\forall l \in \Omega_j$, which involves transmission of $\sum_{l\in\Omega_j} N_l + |\Omega_j|N_j$ numbers.
    \item Distribute $\valpha_j$ and $\phi(\vX_j)^\top\veta_j\vxi_j\vH\ve_l$ to and receive $\valpha_l$ and $\phi(\vX_l)\veta_l\vxi_l\vH\ve_j$ from neighbor $l,\;\forall l \in \Omega_j$, where $2|\Omega_j|N_j + 2\sum_{l\in\Omega_j} N_l$ numbers are transmitted.
\end{itemize}
Thus, when the data are evenly distributed, i.e., $N_1 = N_2 =\cdots = N_J = N$, the communication cost of node $j$ in one ADMM iteration is $\mathcal{O}(|\Omega_j|N)$.
The communication cost is not much because we optimize the communication process, where $\phi(\vX_l)^\top\veta_l\vxi_l\vH\ve_j \in \mathcal{R}^{N_l }$ is transmitted instead of $\phi(\vX_l)^\top\veta_l \in \mathcal{R}^{N_l\times |\Omega_l|}$ according to (\ref{equ: ze}). (Note that $\vxi_l^\top\vH\ve_j$ is a constant matrix such that no extra computation is required when communicating.)

Alg.~\ref{alg: soft} is also computation efficient because each iteration has analytical solution.
In each ADMM iteration, node $j, \forall j$ computes three steps:
\begin{itemize}
    \item Update $\phi(\vX_l)^\top\vz_j, \forall l\in\Omega_j$, of which the computation complexity is less than $$\mathcal{O}\left(|\Omega_j|\left(\sum_{l\in\Omega_j}N_l \right)\left(\max_{l\in\Omega_j}N_l\right)\right).$$
    \item Update $\valpha_j$ and the computation complexity is $\mathcal{O}(N_j^3)$. 
    \item Update $\phi(\vX_j)^\top\veta_j$, of which the computation complexity is $\mathcal{O}(N_j^2|\Omega_j|).$
\end{itemize}
Thus, when the data are evenly distributed, the computation complexity of node $j$ is
$\mathcal{O}\left(\max\{N^3,|\Omega_j|^2 N^2\}\right).$ Compared with central kPCA, where one solves the eigen problem of the global kernel matrix and thus the computation complexity is $\mathcal{O}(J^2N^2)$, Alg.~\ref{alg: soft} is much cheaper since usually $\max_j|\Omega_j|<< J$.
Besides, both the communication and computation cost of Alg.~\ref{alg: soft} are independent with the size of network, which is the property of decentralized algorithms and is better than that of centralized, e.g., server-agent, distributed kPCA.

\section{convergence analysis}\label{sec: convergence}
Different from the original ADMM,  Alg.~\ref{alg: soft} uses another optimization (\ref{equ: U-problem}) to update $\vZ$ due to the implicit mapping $\phi(\cdot)$.
Thus, in this section, we present the convergence analysis for Alg.~\ref{alg: soft}.
We make the following assumption on the hyper-parameter of ADMM.
\begin{assumption}\label{asmp: rho}
The penalty parameter $\rho$ is chosen large enough such that for any $j=1,\cdots,J$, it holds
\begin{equation*}
    \rho \geq \frac{\sqrt{\lambda_1^4(\vK_j)+8|\Omega_j|\lambda_1(\vK_j)\sum_{n=1}^{N_j}(\lambda_n(\vK_j))^3}+\lambda_1^2(\vK_j)}{|\Omega_j|\lambda_1(\vK_j)},
\end{equation*}
where $\lambda_1(\vK_j)$ is {\LShi the largest eigenvalue of $\vK_j$.}
\end{assumption}
Such assumption is common on the convergence analysis for ADMM, e.g., \cite{hong2016convergence}.
Then we have the following result on the convergence of Alg.~\ref{alg: soft}, where we bound the successive difference of the augmented Lagrangian function values.
\begin{theorem}\label{the: convergence}
Suppose Assumption~\ref{asmp: rho} is satisfied. Let $\{\valpha_j^{(t)},$ ${\vZ}^{(t)},$ $\veta_j^{(t)}\}$ is generated by Alg.~\ref{alg: soft} at iteration $t$.
Then we have the following:
\begin{equation*}
    \begin{aligned}
        &\mathcal{L}\left(\valpha_j^{(t+1)}, {\vZ}^{(t+1)}, \veta_j^{(t+1)}\right) - \mathcal{L}\left(\valpha_j^{(t)}, {\vZ}^{(t)}, \veta_j^{(t)}\right)\\
        &\leq -\frac{\rho}{2}\sum_{j=1}^J\left(\left\|\phi(\vX_j)\vK_j^{-1}\phi(\vX_j)^\top(\vZ^{(t+1)}-\vZ^{(t)})\vxi_j\right\|_2^2\right)-\sum_{j=1}^J\left(\frac{c_j}{2}- \frac{4\sum_{n=1}^{N_j}(\lambda_n(\vK_j))^3}{\rho}\right)\left\|\valpha_j^{(t+1)}-{\valpha_j}^{(t)}\right\|_2^2
    \end{aligned}
\end{equation*}
where
\begin{equation*}
    \begin{aligned}
        c_j\geq
        \left\{
        \begin{array}{cc}
            (\rho|\Omega_j| - \lambda_1)\lambda_1, & \mathrm{if}\; \frac{\lambda_1 + \lambda_n}{2|\Omega_j|} \leq \rho, \\
            (\rho|\Omega_j| - \lambda_n)\lambda_n, &\mathrm{if}\; \frac{\lambda_1 + \lambda_n}{2|\Omega_j|} > \rho.
        \end{array}
        \right.
    \end{aligned}
\end{equation*}
\end{theorem}
Theroem~\ref{the: convergence} points out that when the hyper-parameter $\rho$ is chosen sufficent large (e.g., Assumption~\ref{asmp: rho}), the value of augmented Lagrangian function will monotonically decrease.
Note that the augmented Lagrangian function is bounded from below and thus it is convergent.
To proof Theorem~\ref{the: convergence}, we first decompose it into three differences.
The following lemma bounds the difference between $\mathcal{L}(\valpha_j^{(t+1)}, {\vZ}^{(t+1)}, \veta_j^{(t+1)})$ and $\mathcal{L}(\valpha_j^{(t+1)}, {\vZ}^{(t)}, \veta_j^{(t+1)})$.

\begin{lemma}\label{lem: error1}
Let
$$\mathcal{E}_1 \triangleq \mathcal{L}\left(\valpha_j^{(t+1)}, {\vZ}^{(t+1)}, \veta_j^{(t+1)}\right)
        -\mathcal{L}\left(\valpha_j^{(t+1)}, {\vZ}^{(t)}, \veta_j^{(t+1)}\right),$$ we have,
\begin{equation*}
\begin{aligned}
    \mathcal{E}_1&\leq-\frac{\rho}{2}\sum_{j=1}^J\left(\left\|\phi(\vX_j)\vK_j^{-1}\phi(\vX_j)^\top(\vZ^{(t+1)}-\vZ^{(t)})\vxi_j\right\|_2^2\right).
\end{aligned}
\end{equation*}
\end{lemma}

\begin{proof}
From the definition of $\mathcal{U}(\valpha_j^{(t+1)}, \vZ^{(t)}, \veta_j^{(t+1)})$ and $\mathcal{L}(\valpha_j^{(t+1)}, \vZ^{(t)}, \veta_j^{(t+1)})$, we have
\begin{equation*}
    \begin{aligned}
        &\mathcal{U}\left(\valpha_j^{(t+1)}, \vZ^{(t)}, \veta_j^{(t+1)}\right)
        -\mathcal{L}\left(\valpha_j^{(t+1)}, {\vZ}^{(t)},\veta_j^{(t+1)}\right)\\
        &=\frac{\rho}{2}\sum_{j=1}^J\left(\left\|\phi(\vX_j)\valpha_j\vE_j-\vZ\vxi_j\right\|_2^2 - \left\|\phi(\vX_j)\valpha_j\vE_j-\phi(\vX_j)\vK_j^{-1}\phi(\vX_j)^\top\vZ\vxi_j\right\|_2^2 \right)\\
        &=\frac{\rho}{2}\sum_{j=1}^J\left(\left\|\vZ\vxi_j\right\|_2^2-\left\|\phi(\vX_j)\vK_j^{-1}\phi(\vX_j)^\top\vZ\vxi_j\right\|_2^2\right)\\
        &= \frac{\rho}{2}\sum_{j=1}^J\left\|\left(\vI-\phi(\vX_j)\vK_j^{-1}\phi(\vX_j)^\top\right)\vZ^{(t)}\vxi_j\right\|_2^2. \\
    \end{aligned}
\end{equation*}
$\mathcal{U}(\valpha_j, \vZ, \veta_j)$ is $\frac{\rho}{2}$-strong convex with respect to $\vZ\vxi_j$.
Thus, we have
\begin{equation*}
    \begin{aligned}
    & \mathcal{U}\left(\valpha_j^{(t+1)}, {\vZ}^{(t+1)}, \veta_j^{(t+1)}\right)
    -\mathcal{U}\left(\valpha_j^{(t+1)}, {\vZ}^{(t)}, \veta_j^{(t+1)}\right)
    \leq -\frac{\rho}{2}\sum_{j=1}^J\left\|(\vZ^{(t+1)}-{\vZ}^{(t)})\vxi_j\right\|_2^2,
    \end{aligned}
\end{equation*}
because $\vZ^{(t+1)}$ is the optimal solution of problem (\ref{equ: U-problem}).

\begin{equation*}
    \begin{aligned}
        &\mathcal{E}_1 =\mathcal{L}\left(\valpha_j^{(t+1)}, {\vZ}^{(t+1)}, \veta_j^{(t+1)}\right)
        -\mathcal{U}\left(\valpha_j^{(t+1)}, {\vZ}^{(t)}, \veta_j^{(t+1)}\right)\\
        &\quad\quad+\mathcal{U}\left(\valpha_j^{(t+1)}, {\vZ}^{(t)}, \veta_j^{(t+1)}\right)
        -\mathcal{L}\left(\valpha_j^{(t+1)}, {\vZ}^{(t)}, \veta_j^{(t+1)}\right)\\
        &=  \mathcal{L}\left(\valpha_j^{(t+1)}, {\vZ}^{(t+1)}, \veta_j^{(t+1)}\right) -\mathcal{U}\left(\valpha_j^{(t+1)}, {\vZ}^{(t+1)}, \veta_j^{(t+1)}\right)\\
        &\quad\quad+\mathcal{U}\left(\valpha_j^{(t+1)}, {\vZ}^{(t+1)}, \veta_j^{(t+1)}\right)
        -\mathcal{U}\left(\valpha_j^{(t+1)}, {\vZ}^{(t)}, \veta_j^{(t+1)}\right)\\
        &\quad\quad+\mathcal{U}\left(\valpha_j^{(t+1)}, {\vZ}^{(t)}, \veta_j^{(t+1)}\right)
        -\mathcal{L}\left(\valpha_j^{(t+1)}, {\vZ}^{(t)}, \veta_j^{(t+1)}\right)\\
        &\leq
        \frac{\rho}{2}\sum_{j=1}^J\left(-\left\|(\vI-\phi(\vX_j)\vK_j^{-1}\phi(\vX_j)^\top)\vZ^{(t+1)}\vxi_j\right\|_2^2 \right.-\left\|(\vZ^{(t+1)}-\vZ^{(t)})\vxi_j\right\|_2^2\\
        &\quad\quad\quad\quad\quad\quad+\left.\left\|(\vI-\phi(\vX_j)\vK_j^{-1}\phi(\vX_j)^\top)\vZ^{(t)}\vxi_j\right\|_2^2\right)\\
        &\leq \frac{\rho}{2}\sum_{j=1}^J\left(\left\|(\vI-\phi(\vX_j)\vK_j^{-1}\phi(\vX_j)^\top)(\vZ^{(t+1)}-\vZ^{(t)})\vxi_j\right\|_2^2 -\left\|(\vZ^{(t+1)}-\vZ^{(t)})\vxi_j\right\|_2^2\right)\\
        &= -\frac{\rho}{2}\sum_{j=1}^J\left(\left\|\phi(\vX_j)\vK_j^{-1}\phi(\vX_j)^\top(\vZ^{(t+1)}-\vZ^{(t)})\vxi_j\right\|_2^2\right).
     \end{aligned}
\end{equation*}
Then we complete the proof.
\end{proof}

Next, the difference between $\mathcal{L}(\valpha_j^{(t+1)}, {\vZ}^{(t)}, \veta_j^{(t+1)})$ and $\mathcal{L}(\valpha_j^{(t+1)}, {\vZ}^{(t)}, \veta_j^{(t)})$ is bounded by the lemma below.

\begin{lemma}\label{lem: error2}
Let
$$\mathcal{E}_2 \triangleq \mathcal{L}\left(\valpha_j^{(t+1)}, {\vZ}^{(t)}, \veta_j^{(t+1)}\right)
        -\mathcal{L}\left(\valpha_j^{(t+1)}, {\vZ}^{(t)}, \veta_j^{(t)}\right),$$
then $\mathcal{E}_2$ is bounded by
\begin{equation*}
    \begin{aligned}
        \mathcal{E}_2\leq \frac{4}{\rho}\sum_{j=1}^J\left(\sum_{n=1}^{N_j}\left(\lambda_n(\vK_j)\right)^3\left\|\valpha_j^{(t+1)}-\valpha_j^{(t)}\right\|^2_F\right),
    \end{aligned}
\end{equation*}
where $\lambda_n(\vK)$ is the $n$-th eigenvalue of $\vK$.
\end{lemma}
\begin{proof}
From the update of $\valpha_j^{(t+1)}$, we have
\begin{equation*}
    \begin{aligned}
    &\left(2\vK_j^2- \rho|\Omega_j|\vK_j\right)\valpha_j^{(t+1)}=\phi(\vX_j)^\top\left(\veta_j^{(t)}-\rho\vZ ^{(t)}\vxi_j\right)\vE_j^\top.
    \end{aligned}
\end{equation*}
Substituting (\ref{equ: update  phi eta}) to it, we have,
\begin{equation}\label{equ: alpha and eta}
    \begin{aligned}
    &\left(2\vK_j^2- \rho|\Omega_j|\vK_j\right)\valpha_j^{(t+1)}\\ &=\phi(\vX_j)^\top\veta_j^{(t+1)}\vE_j^\top-\rho\vK_j\valpha_j^{(t+1)}\vE_j\vE_j^\top
    +\rho\phi(\vX_j)^\top\vZ^{(t)}\vxi_j\vE_j^\top-\rho\phi(\vX_j)^\top\vZ ^{(t)}\vxi_j\vE_j^\top\\
    &\Longleftrightarrow 2\vK_j^2\valpha_j^{(t+1)} = \phi(\vX_j)^\top\veta_j^{(t+1)}\vE_j^\top
    \end{aligned}
\end{equation}
Therefore, it holds
\begin{equation*}
    \begin{aligned}
        \mathcal{E}_2 &=\mathcal{L}\left(\valpha_j^{(t+1)}, {\vZ}^{(t)}, \veta_j^{(t+1)}\right)
        -\mathcal{L}\left(\valpha_j^{(t+1)}, {\vZ}^{(t)}, \veta_j^{(t)}\right)\\
        &\overset{(\ref{equ: update  phi eta})}= \frac{1}{\rho}\sum_{j=1}^J \mathrm{tr}\left(\left(\veta_j^{(t+1)}-\veta_j^{(t)}\right)^\top\phi(\vX_j)\vK_j^{-1}\phi(\vX_j)^\top\left(\veta_j^{(t+1)}-\veta_j^{(t)}\right)\right) \\
        &= \frac{1}{\rho}\sum_{j=1}^J\left\|\Sigma_j^{-1/2}\vV_j^\top\phi(\vX_j)^\top\left(\veta_j^{(t+1)}-\veta_j^{(t)}\right)\right\|^2_F\\
        &\leq \frac{1}{\rho}\sum_{j=1}^J\left\|\Sigma_j^{-1/2}\vV_j^\top\phi(\vX_j)^\top\left(\veta_j^{(t+1)}-\veta_j^{(t)}\right)\vE_j^\top\right\|^2_F\\
        &\overset{(\ref{equ: alpha and eta})}= \frac{1}{\rho}\sum_{j=1}^J\left\|2\Sigma_j^{-1/2}\vV_j^\top\vK_j^2\left(\valpha_j^{(t+1)}-\valpha_j^{(t)}\right)\right\|^2_F \\
        &= \frac{4}{\rho}\sum_{j=1}^J\left\|\Sigma_j^{3/2}\vV_j^\top\left(\valpha_j^{(t+1)}-\valpha_j^{(t)}\right)\right\|^2_F \\
        &\leq \frac{4}{\rho}\sum_{j=1}^J\left(\sum_{n=1}^{N_j}(\lambda_n(\vK_j))^3\left\|\valpha_j^{(t+1)}-\valpha_j^{(t)}\right\|^2_F\right)
    \end{aligned}
\end{equation*}
where the singular value decomposition of $\vK_j$ is $\vV\Sigma_j\vV_j^\top$ and we complete the proof.
\end{proof}

In a similar way, $\valpha_j^{(t+1)}$ is the optimal solution of problem (\ref{equ: augmented lagrangian}). Therefore, we have
\begin{equation}\label{equ: error3}
    \begin{aligned}
        \mathcal{E}_3 &\triangleq\mathcal{L}\left(\valpha_j^{(t+1)}, {\vZ}^{(t)}, \veta_j^{(t)}\right)
        -\mathcal{L}\left(\valpha_j^{(t)}, {\vZ}^{(t)}, \veta_j^{(t)}) \right)
        \leq -\sum_{j=1}^J\frac{c_j}{2}\left\|\valpha_j^{(t+1)}-{\valpha_j}^{(t)}\right\|_2^2
    \end{aligned}
\end{equation}
because Assumption~\ref{asmp: rho} is satisfied and thus the augmented Lagrangian function is $\frac{c_j}{2}$-convex with respect to $\valpha_j$, where
\begin{equation*}
    \begin{aligned}
        c_j\geq
        \left\{
        \begin{array}{cc}
            (\rho|\Omega_j| - \lambda_1)\lambda_1, & \mathrm{if}\; \frac{\lambda_1 + \lambda_n}{2|\Omega_j|} \leq \rho, \\
            (\rho|\Omega_j| - \lambda_n)\lambda_n, &\mathrm{if}\; \frac{\lambda_1 + \lambda_n}{2|\Omega_j|} > \rho.
        \end{array}
        \right.
    \end{aligned}
\end{equation*}
Finally, we are at the stage of proofing Theorem~\ref{the: convergence}.

\begin{proof}
\begin{equation*}
    \begin{aligned}
        &\mathcal{L}\left(\valpha_j^{(t+1)}, {\vZ}^{(t+1)}, \veta_j^{(t+1)}\right) - \mathcal{L}\left(\valpha_j^{(t)}, {\vZ}^{(t)}, \veta_j^{(t)}\right)\\
        &= \mathcal{L}\left(\valpha_j^{(t+1)}, {\vZ}^{(t+1)}, \veta_j^{(t+1)}\right)
        -\mathcal{L}\left(\valpha_j^{(t+1)}, {\vZ}^{(t)}, \veta_j^{(t+1)}\right)\\
        &\quad+\mathcal{L}\left(\valpha_j^{(t+1)}, {\vZ}^{(t)}, \veta_j^{(t+1)}\right)
        -\mathcal{L}\left(\valpha_j^{(t+1)}, {\vZ}^{(t)}, \veta_j^{(t)}\right) \\
        &\quad+\mathcal{L}\left(\valpha_j^{(t+1)}, {\vZ}^{(t)}, \veta_j^{(t)}\right)
        -\mathcal{L}\left(\valpha_j^{(t)}, {\vZ}^{(t)}, \veta_j^{(t)}\right) \\
        &=\mathcal{E}_1 + \mathcal{E}_2 +\mathcal{E}_3,
    \end{aligned}
\end{equation*}
where
\begin{equation*}
    \begin{aligned}
        \mathcal{E}_1 &= \mathcal{L}\left(\valpha_j^{(t+1)}, {\vZ}^{(t+1)}, \veta_j^{(t+1)}\right)
        -\mathcal{L}\left(\valpha_j^{(t+1)}, {\vZ}^{(t)}, \veta_j^{(t+1)}\right)\\
        \mathcal{E}_2 &= \mathcal{L}\left(\valpha_j^{(t+1)}, {\vZ}^{(t)}, \veta_j^{(t+1)}\right)
        -\mathcal{L}\left(\valpha_j^{(t+1)}, {\vZ}^{(t)}, \veta_j^{(t)}\right)\\
        \mathcal{E}_3 &=\mathcal{L}\left(\valpha_j^{(t+1)}, {\vZ}^{(t)}, \veta_j^{(t)}\right)
        -\mathcal{L}\left(\valpha_j^{(t)}, {\vZ}^{(t)}, \veta_j^{(t)}\right).
    \end{aligned}
\end{equation*}
Then, combining the results in Lemma~\ref{lem: error1}, Lemma~\ref{lem: error2} and (\ref{equ: error3}), we obtain Theorem~\ref{the: convergence} and complete the proof.
\end{proof}

\section{Experiments}\label{sec: exp}

In this section, we present experimental result to show the performance of Alg.~\ref{alg: soft} on both accuracy and efficiency.

\subsection{Experimental Setting}
\textbf{Metric.}
In experiments, we care about two aspects: solution quality and running time. Since this is the first algorithm for decentralized kPCA, we evaluate the quality by comparing with the central method. 
We define the following two solutions:
\begin{itemize}
    \item $\valpha_j\in\mathcal{R}^{N_j}$: the solution of Alg.~\ref{alg: soft}.
    \item $\valpha_{\mathrm{gt}}\in\mathcal{R}^{N}$: the solution of central kPCA, which is regarded as the ground truth. That is the eigenvector of global kernel matrix $\vK$ that associated with the largest eigenvalue.
\end{itemize}
Then the similarity of any $\vw_j = \phi(\vX_j)\valpha_j$ 
to the global solution $\vw_{\mathrm{gt}} = \phi(\vX)\valpha_{\mathrm{gt}}$ cab be calculated as follow,
\begin{equation*}
\begin{aligned}
    &\mathrm{Similarity} =
    \frac{\vw_j^\top \vw}{\|\vw_j\|\|\vw\|}
    =\frac{\valpha_j\mathcal{K}(\vX_j, \vX)\valpha_{\mathrm{gt}}}{\sqrt{\|\valpha_j^\top\vK_j\valpha_j\|\|\valpha_{\mathrm{gt}}^\top\vK\valpha_{\mathrm{gt}}}\|}.
\end{aligned}
\end{equation*}

Running time is another aspect we care about. 
We compare the running time of Alg.~\ref{alg: soft} and the central kPCA that obtains $\valpha_{\mathrm{gt}}$. For central kPCA, all data are collected into one node to compute the global Gram kernel matrix, and then SVD is performed on it.


\textbf{Hardware and software.}
Both Alg. 1 and the central kPCA are implemented 
on a truly parallel architecture by Python with the package MPI4PY.
All the experiments are conducted on the $\pi\;2.0$ cluster supported by the Center for High Performance Computing (HPC) at Shanghai Jiao Tong University, where each server has two Intel(R) Xeon(R) Gold 6248 CPUs (2.5GHz, 20 cores) and $40$G memory.
In servers that we applied for from HPC, all CPUs and memory are for exclusive use.
Recall that we use cores in CPUs to simulate nodes in network.
In the following, we use one CPU to simulate $20$ network nodes out of consideration of computational efficiency.
All the experiments are repeated $100$ times.
Python codes of Alg.~\ref{alg: soft} and following experiments are available in \url{https://github.com/Yruikk/DKPCA-ADMM}.

\textbf{Dataset and data preprocessing.}
We use MNIST dataset \cite{deng2012mnist}, which contain images of $28\times 28$ pixels, from digit $0$ to digit $9$.
We vectorize each image to a $784\times 1$ vector.
Due to the high computational complexity of central kPCA, we cannot use entire data of MNIST. Thus, we follow the setting in \cite{forero2010consensus, yang2021achieving}:
images for digits $0,3,5,8$ are used and they are randomly and evenly distributed to local nodes.

When performing central kPCA or Alg.~\ref{alg: soft}, both local and global kernels are centralized by the following formulation.
\begin{equation*}
    \vK_c = \vK - \frac{1}{m}{\bf 1}_m\vK - \frac{1}{n}\vK{\bf 1}_n + \frac{1}{mn}{\bf 1}_m\vK{\bf 1}_n,
\end{equation*}
where $\vK\in\mathcal{R}^{m\times n}$ and $\vK_c$ is its centralized version.
${\bf 1}_n$ is a $n\times n$ matrix with all elements equaling to $1$.

\textbf{Tuning the hyper-parameter for ADMM.}
For ADMM framework, there is a hyper-parameter for each constraint. When introducing and analyzing Alg. 1, we use the same and fixed $\rho$ for all the constraints, for easy representation and understanding. In experiments, we use the following tuning strategy, which is of course not optimal but works well in our experiments. 

Let $\rho^{(1)}$ correspond to the constraint $\phi(\vX_j)\valpha_j =  \phi(\vX_j)\vK_j^{-1} \phi(\vX_j)^\top\vz_j$ and  $\rho^{(2)}$ to constraint $\phi(\vX_j)\valpha_j =  \phi(\vX_j)\vK_j^{-1} \phi(\vX_j)^\top\vz_p ,\forall p \in \Omega_j, \forall j = 1,\cdots,J$.
We set $\rho^{(1)} = 100$ but initially use a relatively small $\rho^{(2)} = 10$ and then gradually increase $\rho^{(2)}$ to $50$ and $100$. 


\subsection{Experimental Results}
\begin{figure}[h]
\begin{center}
\centerline{\includegraphics[width=0.35\textwidth]{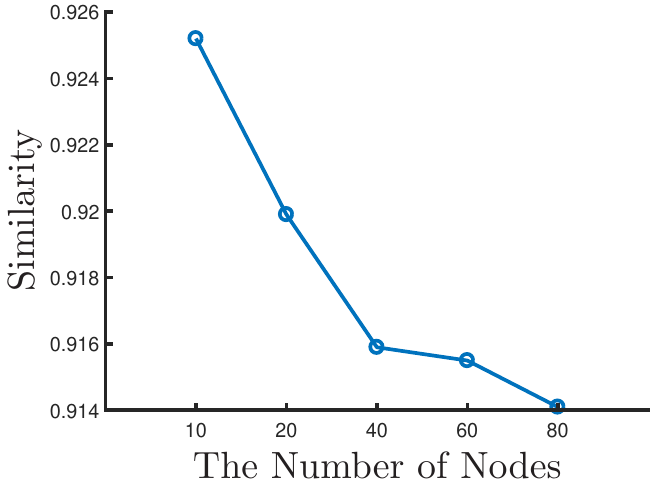}
\includegraphics[width=0.35\textwidth]{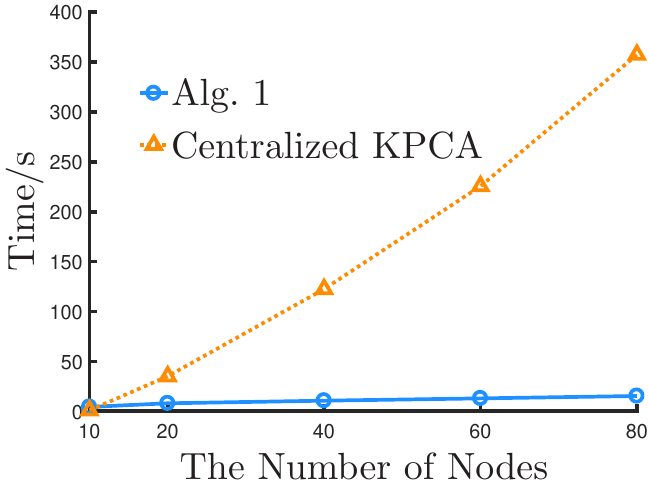}
}
\caption{The average similarity of $\valpha_j$ with respect to the number of network nodes. Each node has $100$ images from MNIST and communicates with $4$ neighbor nodes closest to it.}
\label{fig: exp-1}
\end{center}
\end{figure}

Consider a situation, where each network node holds $100$ images and communicates with $4$ neighbors closest to it.
As a distributed method,  Alg.~\ref{alg: soft} is expected to be faster than a central algorithm and the similarities between the directions given by $\valpha_i$ and $\valpha_{\mathrm{gt}}$ are displayed in Fig. \ref{fig: exp-1}, which includes the performance with different numbers of nodes. Overall, the similarity is very high, even when the network is large. For example, when there are 80 nodes, the similarity is kept above 0.912 and the speed advantage is obvious, which coincides with our previous evaluation of computation complexity, where that of central kPCA is $\mathcal{O}(J^2N^2)$ and that of Alg.~1 is independent from the network size.



\begin{figure}[t]
\begin{center}
\centerline{\includegraphics[width=0.35\textwidth]{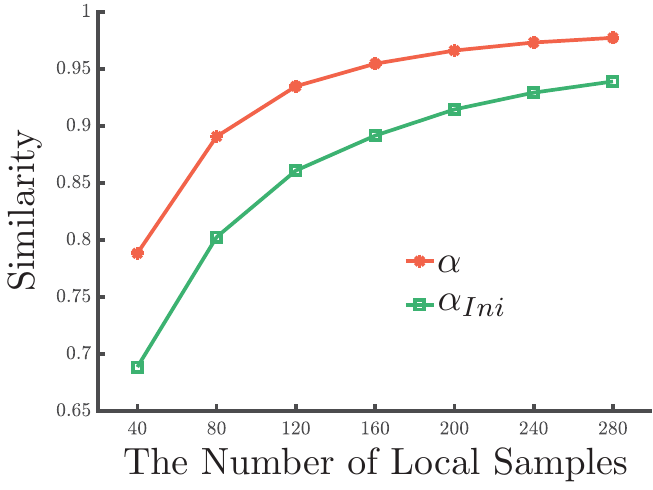}
}
\caption{The average similarity of $\valpha_j$ and $(\valpha_j)_{\mathrm{local}}$ with respect to the number of local samples in a $20$-node network. For each node, there are $4$ 
neighbors.}
\label{fig: exp-2}
\end{center}
\end{figure}

Before, the number of samples $N_j$ in each node is kept unchanged. Now we vary  $N_j$ from $40$ to $300$ in a $20$-node network and set the solution of kPCA on local data, denoted by $(\valpha_j)_{\mathrm{local}}$, as the baseline. The corresponding results are presented in Fig.~\ref{fig: exp-2}.
When there are only a few local data, the similarity of $(\valpha_j)_{\mathrm{local}}$ 
is low and Alg. 1 could improve the performance by adding consensus constraints to make nodes cooperated. Not surprisingly, the improvement is less significant when there are more data in local nodes. 

\begin{figure}[t]
\begin{center}
\centerline{\includegraphics[width=0.35\textwidth]{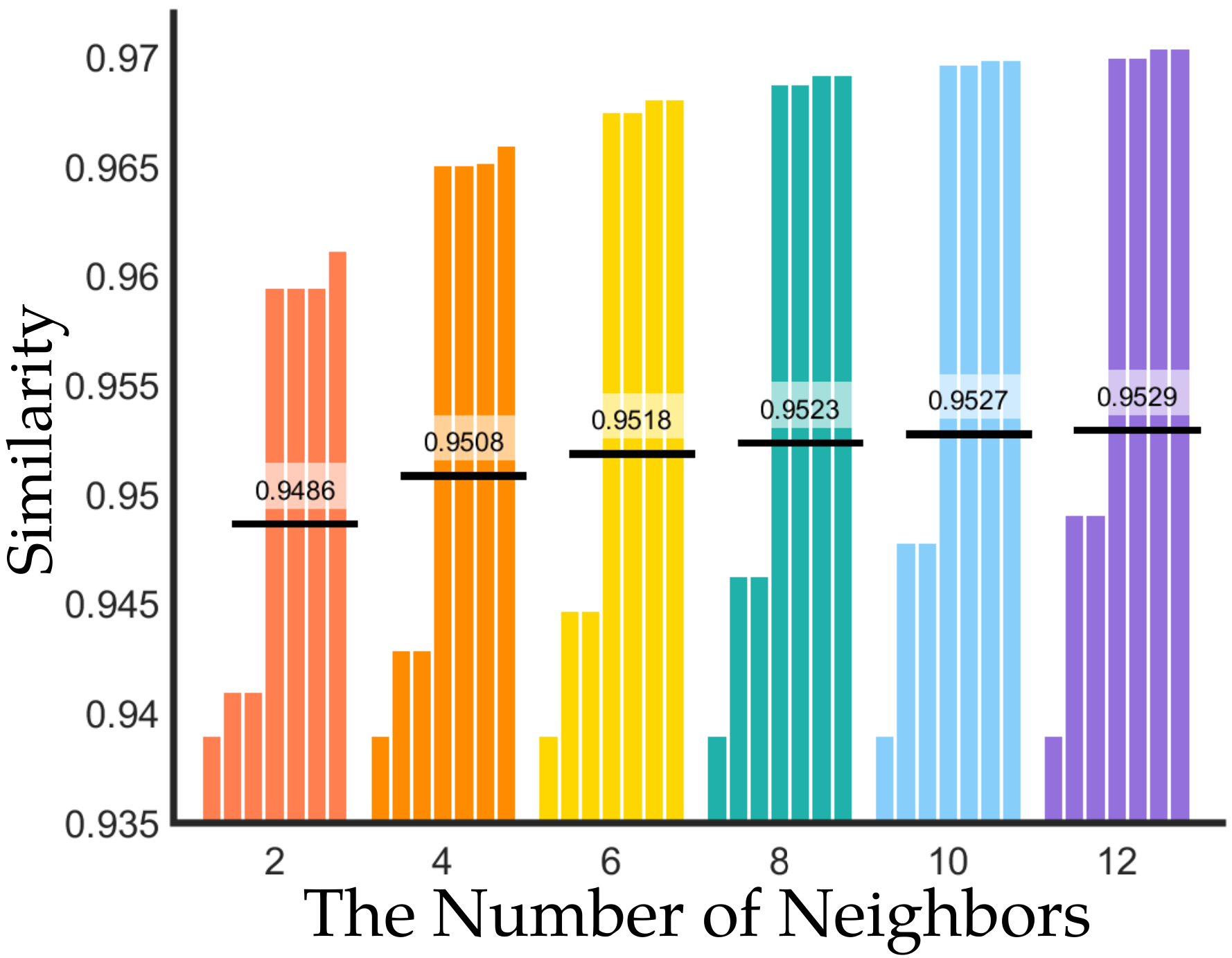}
}
\caption{The average similarity of $\valpha_j$ and $(\valpha_j)_{\mathrm{Nei}}$ with respect to the number of neighbors in a $20$-node network. Each node has $100$ images from MNIST. The histogram shows the similarity of $\valpha_j$ after each iteration of Alg.~\ref{alg: soft}. The black solid line is the average similarity of $(\valpha)_{\mathrm{Nei}}$.}
\label{fig: exp-3}
\end{center}
\end{figure}

Last we conduct experiments for different numbers of neighbors. The total number of nodes is 20 and each node contains $100$ images from MNIST. The numbers of neighbors are set as $|\Omega_j|=\{2,\;4,\;6,\;8,\;10,\;12\}$ and the corresponding results are displayed in Fig. \ref{fig: exp-3}. In Alg. 1, the information is cooperated by the consensus constraint gradually. Accordingly, one may observe the similarity is increasing with more update iterations, showing the information diffusion process. As a comparison, we can directly take all the data in neighbors and calculate. The solution is denoted as  $(\valpha_j)_{\mathrm{Nei}}$ and the average similarity is given in Fig. 5 as the black solid lines. For all the situations, with about 4 iterations, the similarity given by Alg. 1 exceeds that of obtaining all neighbor data. When Alg.~\ref{alg: soft} converges, the similarity is similar to ($|\Omega_j|=2$) or much better than ($|\Omega_j|=\{4,\;6,\;8,\;10,\;12\}$) that of gathering $12$ neighboring nodes, because the projection consensus constraints effectively cooperate data from all the nodes.

\section{Conclusion}\label{sec: conclusion}
This paper derived a decentralized framework for kPCA, where a novel projection consensus constraint is proposed to decouple the original optimization.
The proposed projection consensus constraint is based on the optimal approximation, where we minimized the Euclidean distance between local solutions and the global optimum.
Different from the traditional consensus constraint, the projection consensus constraint can deal with not only data heterogeneous but the representation discrepancy.
A fast, fully non-parametric solving algorithm based on ADMM was then proposed, where each iteration has analytic solution.
In order to deal with implicit mapping, one of the ADMM iteration is solved by approximation.
Therefore, we provided a convergence analysis, showing the algorithm still converges as long as the hyper-parameter of ADMM is sufficient large.

The proposed algorithm was then implemented on a truly parallel architecture based on MPI.
Experimental results on MNIST demonstrate the effectiveness and the efficiency of the proposed algorithm.
Compared with central kPCA, the running time of proposed decentralized algorithm is much less and is independent from the network size.
Thus, it is more suitable for applications with large dataset and large-scale network.
For future works, it is worth trying the application of random features such that raw data exchange is no longer required.


%

\ifCLASSOPTIONcaptionsoff
  \newpage
\fi



\bibliographystyle{IEEEtran}
%

\bibliography{thesis}

%

\end{document}